\documentclass{article}
\usepackage{graphicx} 
\usepackage{tensor}
\usepackage{amssymb,amsmath,mathrsfs,amsthm, hyperref}
\usepackage{mathtools}
\usepackage[usenames,dvipsnames]{xcolor}

\usepackage[textsize=tiny]{todonotes}

\hypersetup{
	colorlinks=true,            
        breaklinks=true,
	linkcolor=Black,
	citecolor=Blue,
	urlcolor=Blue            
 }

\usepackage{graphicx, tikz}
\usetikzlibrary{cd}
\usepackage[authordate,maxcitenames=3,dashed=false,backend=biber,natbib=true]{biblatex-chicago}

\newtheorem{prop}{Proposition}
\newtheorem{corollary}{Corollary}[prop]

\title{On the geometric trinity of gravity, non-relativistic limits, and Maxwell gravitation}
 \author{Eleanor March\footnote{Faculty of Philosophy, University of Oxford, UK. eleanor.march@philosophy.ox.ac.uk},~ William J.~Wolf\footnote{Faculty of Philosophy, University of Oxford, UK. william.wolf@philosophy.ox.ac.uk}, \& James Read\footnote{Faculty of Philosophy, University of Oxford, UK. james.read@philosophy.ox.ac.uk}}
\date{}

\begin{document}

\maketitle

\begin{abstract}
   We show that the dynamical common core of the recently-discovered non-relativistic geometric trinity of gravity is Maxwell gravitation. Moreover, we explain why no analogous distinct dynamical common core exists in the case of the better-known relativistic geometric trinity of gravity.
\end{abstract}

\tableofcontents

\section{Introduction}\label{sec:intro}

\emph{Inter alia}, the following questions surely count as mainstream in contemporary philosophy of spacetime physics:
\begin{enumerate}
    \item What is the `correct' spacetime setting for Newtonian gravity, especially in light of Newton's Corollary VI?\footnote{\label{fn-vi}Recall that Newton's Corollary VI reads as follows: ``If bodies are moving in any way whatsoever with respect to one another and are urged by equal accelerative forces along parallel lines, they will all continue to move with respect to one another in the same way as they would if they were not acted on by those forces.'' \parencite[p.~99]{Newton}.} (On this topic, see e.g.~\textcite{Dewar, Knox2014-KNONSS, TehNG, Wallace3, Weatherall4, Weatherall7}.)
    \item Are there spacetime theories which are in some sense or other `equivalent' to general relativity, and what would be the philosophical significance of such theories, were they to exist? (On this topic, see e.g.~\textcite{WolfReadSanchioni, WolfRead, Knox2011-KNONTA, DurrRead, Rosenstock2015-ROSOEA-2, BAIN200637}.)
    \item How is one  to take the non-relativistic limit of general relativity, and what is the resulting theory? (On this topic, see e.g.~\textcite{Fletcher2019-FLEOTR-3, Malament1986, 1976_Kunzle}.)
\end{enumerate}
Until now, discussions of these questions have, broadly speaking, been isolated from one another. Our purpose in this article is to show that these questions (and answers to said questions) are in fact related to one another in intimate and significant ways.

To explain what we mean here, begin with question (1). (For the time being a qualitative account will suffice; the mathematics to substantiate the claims made here will follow later in this article.) Typically, Newtonian gravitation theory (NGT) in its potential-based formulation \emph{chez} Laplace and Poisson is taken to be set in a flat spacetime; gravitational effects in this spacetime are encoded in the gravitation potential which leads to test bodies not traversing geodesics of the flat, compatible connection. This being said, NGT has a hidden symmetry (sometimes referred to as `Trautman symmetry' \parencite{TehNG}): if one (a) subjects \emph{all} material bodies to an additional constant gravitational field, and (b) changes one's standard of straightness (i.e.,~one's derivative operator) to compensate for this, then in fact no physical change ensues. (This is related to Newton's Corollary VI, as we will explain below; cf.~\textcite{ReadTeh2}.) When one moves to a new formalism purged of this additional symmetry,\footnote{Here, we set aside the differences between what are known as `reduction' and `internal sophistication' about symmetries: see \textcite{Dewar, MartensRead}.} one arrives at the structure of Newton-Cartan theory (NCT): a non-relativistic spacetime theory in which gravitational effects are---just as in the case of general relativity (GR)---manifestations of spacetime curvature.

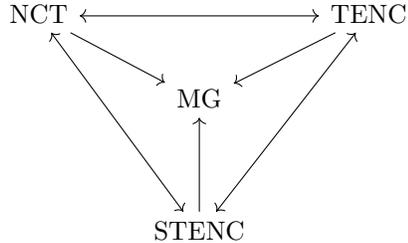
\begin{figure}
\begin{center}
\begin{tikzcd}
\text{NCT} \arrow[rr, leftrightarrow] \arrow[dr, rightarrow] \arrow[dddr, leftrightarrow]
& & \text{TENC} \arrow[dl, rightarrow] \arrow[dddl, leftrightarrow] \\
& \text{MG} \arrow[dd, leftarrow] & \\
& & & \\
& \text{STENC} &
\end{tikzcd}
\end{center}
\caption{Maxwell gravitation as the common core of the non-relativistic geometric trinity of gravity.}
\label{fig:2}
\end{figure}

This much is well-known. But there remains some ambiguity in the literature as to how NGT relates to another spacetime theory known as `Maxwell gravitation' (MG), also developed in light of Newton's Corollary VI. (On this theory, see \textcite{Saunders3, Dewar, Chen, March, march2}.) Moreover, recently NGT has been shown to in fact admit of an interpretation whereby it is a theory with a \emph{torsionful} geometry \parencite{ReadTeh, Schwartz_2023}, in the sense that the gravitational potential can be associated with the torsion of the `mass gauge field' which arises when one gauges the Bargmann algebra (for more on the mass gauge field, see \textcite{Andringa:2010it, TehNG, ReadTeh, Wolf:2021ydy}). We shall refer to this torsional interpretation of NGT as TENC, or the `teleparallel equivalent of Newton-Cartan theory', for reasons which will become apparent shortly. Even less well-known (indeed, we might say, almost unknown!) is that both TENC and NCT are equivalent to an alternative non-relativistic theory, recently dubbed the `symmetric teleparallel equivalent of Newton-Cartan theory' (STENC), in which gravitational effects are manifestations neither of curvature (as in NCT) nor of torsion (as in TENC), but of spacetime non-metricity \parencite{Wolf:2023rad}.\footnote{To remind the reader: `curvature' quantifies the extent to which parallel transport of a vector along a closed loop doesn't preserved angles; `torsion' quantifies the extent to which parallel transport in two directions doesn't commute; `non-metricity' quantifies the extent to which parallel transport of a vector along a closed loop doesn't preserve the length of that vector. For further background, see e.g.~\textcite{Hehl}.} Here, we demonstrate that these pieces fit together in the following way: NCT, TENC and STENC constitute a `non-relativistic geometry trinity'; the structure common to all said theories (the `common core', in the sense of \textcite{BLBJR}) \emph{just is} the structure of MG (see Figure \ref{fig:2}).\footnote{The same notion of a common core is also discussed in e.g.~\textcite{DeHaroButterfield}.}

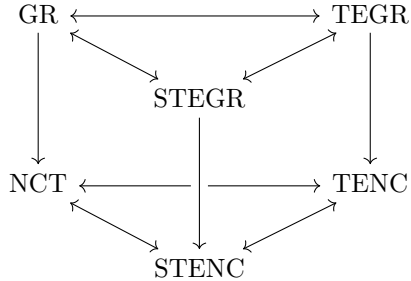
\begin{figure}
\begin{center}
\begin{tikzcd}
\text{GR} \arrow[rr, leftrightarrow] \arrow[dr, leftrightarrow] \arrow[dd]
& & \text{TEGR} \arrow[dl, leftrightarrow] \arrow[dd] \\
& \text{STEGR}  & \\
\text{NCT} \arrow[rr, leftrightarrow] \arrow[dr, leftrightarrow]
& & \text{TENC} \arrow[dl, leftrightarrow] \\
& \text{STENC} \arrow[uu, leftarrow, crossing over] &
\end{tikzcd}
\end{center}
\caption{The relativistic geometric trinity and the non-relativistic geometric trinity as its non-relativistic limit.}
\label{fig:1}
\end{figure}

Already, this illuminates quite substantially the connections between these four non-relativistic theories of spacetime and gravity. And yet, that is only the beginning of the story. Taking now together questions (2) and (3) in our above list, it is becoming increasingly well-known to philosophers of physics that there exists a `geometric trinity' of relativistic theories of gravitation, of which GR constitutes but one node (see e.g.~\textcite{Heisenberg, Capozziello:2022zzh} for recent reviews in the physics literature). The other two nodes are `teleparallel gravity' (TEGR), in which gravitational effects are a manifestation of exclusively spacetime torsion, and `symmetric telelparallel gravity' (STEGR), in which gravitational effects are a manifestation of exclusively spacetime non-metricity. Recently, \textcite{Wolf:2023rad} have shown that the above-discussed non-relativistic trinity (\emph{sans} any mention of MG) is indeed the non-relativistic limit (in the sense of a $1/c^2$ expansion \emph{\`{a} la} \textcite{Schwartz_2023}) of this relativistic geometry trinity. More specifically, the torsional non-relativistic theory TENC arises when one takes the non-relativistic limit of the torsional relativistic theory TEGR, while the non-metric non-relativistic theory STENC arises as the non-relativistic limit of non-metric relativistic theory STEGR. The web of connections is, therefore, as presented in Figure \ref{fig:1} (in that figure, for clarity, we omit MG).

What we add to this discussion in the present paper is an answer to the following question: does there exist a `common core' of the relativistic geometric trinity in the same sense that MG is the common core of the non-relativistic trinity, and if so is it the case that MG is the non-relativistic limit of said relativistic common core? Although we address both parts of this question, we should be clear that our answer to the first part is partly in the negative: there is \emph{not} a dynamically distinct common core of the relativistic geometric trinity in the same sense that MG is the dynamically distinct common core of the non-relativistic trinity.

To elaborate: we concur with the recent verdict of \textcite{WolfReadSanchioni}, who argue that the common core of all three of GR, TEGR, and STGR simply is just GR. This is because one can always use the metric to build GR's defining affine structure, as the Levi-Civita connection is the \emph{unique} torsion-free, metric compatible derivative operator; using this derivative operator one can then write down the Einstein equation as usual.
In the non-relativistic case there is again a common core---this time, its kinematics include what has been dubbed in the recent philosophical literature a `standard of rotation' (see \textcite{Weatherall7})---; however---and quite differently to the relativistic case!---that common core leads to dynamics which are \emph{distinct} from those of the the geometric trinity from which they arise.
Later in this article, we will explain exactly how this asymmetry between the relativistic and non-relativistic cases.

To summarise, then, in this article we (a) identify MG as the dynamical common core of the recently-discovered non-relativistic geometric trinity of gravity, and (b) explain how it can be  that no analogous common core exists in the case of the relativistic geometric trinity of gravity. In so doing, we (i) clarify questions in (1) regarding the `correct' spacetime setting for Newtonian gravity, (ii) connect that entire literature up to the geometric trinity of gravity and its Newtonian limit, which has also aroused recent philosophical interest.

More specifically, the structure of the article is this. In \S\ref{sec:trinities}, we remind the reader of the mathematical details of both the relativistic geometric trinity and the non-relativistic geometric trinity. In \S\ref{sec:Maxwell}, we present MG as the common core of the non-relativistic trinity, and connect our discussion to that of the `correct' spacetime setting for Newtonian gravity. In \S\ref{sec:relcommoncore}, we address the matter of the existence (or otherwise) of a relativistic common core. We close in \S\ref{sec:conc}.

\section{Geometric trinities}\label{sec:trinities}

In this section, we recall the mathematical details underlying the existence of the relativistic geometric trinity of gravitation theories (\S\ref{sec:relGT}) and the non-relativistic geometric trinity of gravitation theories (\S\ref{sec:nonrelGT}).

\subsection{The relativistic geometric trinity}\label{sec:relGT}

The `geometric trinity' of gravity refers to a family of three relativistic theories of gravitation: general relativity (GR), the `teleparallel equivalent to general relativity' (TEGR), and the `symmetric teleparallel equivalent to general relativity' (STEGR). These theories are all `equivalent' to each other in the sense that they share equivalent dynamical equations of motion, but distinct in the sense that these shared dynamics result from entirely different geometric degrees of freedom that manifest in each respective theory (see e.g.\ \textcite{Heisenberg, Capozziello:2022zzh, Heisenberg:2018vsk}).  

Kinematical possibilities of general relativity are typically presented as tuples of the form $\langle M, g_{ab}, \Phi \rangle$, where $M$ is a four-dimensional differentiable manifold, $g_{ab}$ is a Lorentzian metric field on $M$, and $\Phi$ represents material fields. The dynamical possibilities of the theory are encoded by the Einstein  equation, which governs the behavior of these spacetime and material fields.
However, the geometric degrees of freedom responsible for sourcing the dynamics of the respective theories in the geometric trinity are properties of the affine connection. We will thus take GR to be a theory given by models of the form $\langle M, g_{ab}, \overset{c}{\nabla}, \Phi \rangle$, where $\overset{c}{\nabla}$ refers to the familiar Levi-Civita derivative operator with non-vanishing \textit{curvature}. Typically, $\overset{c}{\nabla}$ is not included explicitly in the models of GR, for it is fixed uniquely by $g_{ab}$.

Spacetime curvature is defined by
\begin{equation}
R\indices{^a_{b c d}} \xi^b := -2 \nabla_{[c} \nabla_{d]} \xi^a,
\end{equation}
where $\xi^a$ is a smooth vector field. However, curvature is not the only geometric property that a connection can manifest. An affine connection can also possess \textit{torsion} or \textit{non-metricity}. The torsion tensor is given by 
\begin{equation}
T\indices{^c_{a b}} \nabla_c \alpha := 2 \nabla_{[a} \nabla_{b]} \alpha,
\end{equation}
where $\alpha$ is a smooth scalar field; torsion thereby encodes the antisymmetry of a connection. Non-metricity is given by the non-vanishing of the covariant derivative of the metric tensor
\begin{equation}
Q_{a b c} := \nabla_a g_{b c}.
\end{equation}
Heuristically, curvature measures the rotation of a vector when it is parallel transported along a closed curve, torsion measures of the non-closure of the parallelogram formed by two vectors being parallel transported along each other, and non-metricity measures how the length of a vector changes when parallel transported (see e.g.~Figure 1 in \textcite{Heisenberg} or \cite{Hehl}). Note that while curvature and torsion are properties intrinsic to a connection, non-metricity is a relational property between a connection and a metric.

The Levi-Civita connection of GR is special in the sense that it is the unique derivative operator which is both torsion-free (i.e.\ $T\indices{^a_{b c}}=0$) and metric-compatible (i.e.\ $Q_{abc} = 0$), but with generically non-vanishing curvature (i.e.\ $R\indices{^a_{bcd}} \neq 0$). However, in order to build a viable relativistic spacetime theory, it is not necessary to use $\overset{c}{\nabla}$. Indeed, one can decompose a general affine connection as 
\begin{equation}\label{decomp}
    \nabla = (\overset{c}{\nabla}, K\indices{^{a}_{bc}} + L\indices{^{a}_{bc}} ),
\end{equation}
where $K\indices{^{a}_{bc}}$ is known as the `contorsion tensor' and $L\indices{^{a}_{bc}}$ is known as the `distorsion tensor' (here, we use the notation of \textcite[p.~53]{Malament}). The contorsion tensor can be understood as the difference tensor between the Levi-Civita connection and the torsionful (but flat and metric-compatible) connection of TEGR. The disorsion tensor can be understood as the difference tensor between the Levi-Civita connection and non-metric (but flat and torsionless) connection of STEGR.

If---as above---we take GR to be a theory with kinematical possibilities of the form $\langle M, g, \overset{c}{\nabla}, \Phi \rangle$, then TEGR can be taken to be a theory with kinematical possibilities given by $\langle M, g, \overset{t}{\nabla}, \Phi \rangle$, where $\overset{t}{\nabla} = (\overset{c}{\nabla}, K\indices{^{a}_{bc}} )$ refers to the TEGR connection with vanishing curvature and non-metricity, but non-vanishing torsion. Likewise, STEGR is a theory with kinematical possibilities given by $\langle M, g, \overset{n}{\nabla}, \Phi \rangle$, where $\overset{n}{\nabla} = (\overset{c}{\nabla}, L\indices{^{a}_{bc}} )$ refers to the STEGR connection with vanishing curvature and torsion, but non-vanishing non-metricity. 

One can use (\ref{decomp}) as a dictionary by which to translate between these theories. That is, one can rewrite the geometric objects of interest in one theory in terms of the geometric objects of one of the other trinity theories, and thereby witness their equivalence. For example, one can take the curvature tensor for a generic affine connection and use (\ref{decomp}) to relate the curvature of the Levi-Civita connection to that of the TEGR connection (and the associated contorsion tensor), or to that of the STEGR connection (and the associated distorsion tensor). One then finds that the geometric scalar quantities have the following relationship:
\begin{equation}\label{scalars}
    - R = T + B_{T} = Q + B_{Q},
\end{equation}
where $R$ is the curvature scalar, $T$ is the torsion scalar, $Q$ is the non-metricity scalar, and $B_{T/Q}$ refers to boundary terms of the respective theories \parencite{Heisenberg}.
This also illustrates that these theories are dynamically equivalent, as the Lagrangian expressions for all of these theories can be written using the geometric scalars (in the case of GR, recall the Einstein-Hilbert action). Upon utilizing standard variational procedures, the boundary terms that arise in (\ref{scalars}) vanish, resulting in the Einstein equation for all theories (but of course expressed in their particular geometric languages).\footnote{For more on the significance of these boundary terms, see \textcite{WolfRead} for philosophical discussion on concerning their implications for theory equivalence and theory structure, and see \textcite{Oshita:2017nhn} for further physics discussion.}

\subsection{The non-relativistic geometric trinity}\label{sec:nonrelGT}

It was shown recently by \textcite{Wolf:2023rad} that there is a non-relativistic analogue of the geometric trinity, whereby standard Newtonian gravity can likewise be reconceptualised and/or reformulated as a theory of curvature, torsion, or non-metricity. This non-relativistic trinity is obtained by taking the non-relativistic limit of the relativistic trinity. That is, the respective relativistic theories are expanded in powers of $\lambda \coloneqq 1/c^2$. The three nodes of the non-relativistic geometric trinity of gravity are `Newton-Cartan theory' (NCT), the `teleparallel equivalent of Newton-Cartan theory' (TENC), and the `symmetric teleparallel equivalent of Newton-Cartan theory' (STENC).

Following the presentation of the relativistic theories above, we take NCT to be a theory with kinematical possibilities of the form $\langle M, t_a, h^{ab}, \overset{c}{\nabla}, \Phi \rangle$. As is by now well known, this theory emerges as the non-relativistic limit of GR \parencite{1976_Kunzle}. As before, $M$ is a four-dimensional differentiable manifold, $\Phi$ represents material fields, and $\overset{c}{\nabla}$ is a torsion-free and compatible (now in the sense that $\nabla_a t_b = \nabla_a h^{bc} = 0$) derivative operator with non-vanishing curvature. Some important features of NCT are as follows: 
\begin{enumerate}
    \item The metrical structure of non-relativistic theories is notably different from that of relativistic theories, because $t_a$ and $h^{ab}$ refer to degenerate temporal and spatial metrics on $M$: see \textcite[Ch.~4]{Malament}.\footnote{Through this article, we assume temporal orientability, in the sense of \textcite[ch.~4]{Malament}.} Metric compatibility applies separately to both metrics; in addition, $t_a$ and $h^{ab}$ are orthogonal to each other, so that $t_a h^{ab} =0$. Loosely speaking, $t_a$ is supposed to represent Newtonian absolute time, and $h^{bc}$ is supposed to represent Newtonian absolute space.
    \item This NCT connection is given by the following:\footnote{In abstract indices, `$\partial_a$' denotes a coordinate derivative operator, in the sense of \textcite[Ch.\ 1]{Malament}.}
    \begin{align}\label{eq_Galilean_connection}
    \hat\Gamma^a{}_{bc} = \xi^a \partial_{(b} t_{c)} + h^{as}\left(\partial_{(b} \hat{h}_{c)s} - \frac{1}{2} \partial_{s} \hat{h}_{bc} \right) +  h^{an}t_{(b}f_{c)n},
    \end{align}
    Here, $h$ is the spatial metric, $t$ is the temporal metric, $\hat{h}$ is the spatial projector orthogonal to a choice of timelike vector field $\xi$ (i.e., $\xi^a \hat{h}_{ab} \coloneqq 0$), $\phi$ is a scalar field defined from the so-called Newton-Coriolis two-form $f_{ab}=t_{[a}\phi_{b]}$ (i.e., the compatibility conditions do not uniquely single out a connection in contrast with the relativistic connection and this represents a further freedom in choice of connection). See, e.g., \textcite{Malament, Wolf:2023rad, Schwartz_2023, 1976_Kunzle} for some further discussion of Newton-Cartan connections and these objects.
    \item The dynamical possibilities for NCT are encoded in the `geometrised Poisson equation': 
    \begin{equation}\label{geomPoisson}
        R_{ab} = 4 \pi  \rho t_{a} t_{b},
    \end{equation}
    where $R_{ab}$ is the Ricci curvature of the NCT connection $\overset{c}{\nabla}$; moreover, one typically includes the following curvature conditions in one's presentation of NCT (we will discuss these curvature conditions more later in the article\footnote{See also \textcite{Malament, TehNG} for further discussion of the physical significance and meaning of these conditions.}):
    \begin{align}
    R\indices{^{a}_{b}^{c}_{d}} &= R\indices{^{c}_{d}^{a}_{b}}, \label{trautman1} \\ 
    R\indices{^{ab}_{cd}} &= 0. \label{trautman2}
    \end{align}
\end{enumerate}

Moving beyond curvature-based theories, we are now interested in investigating the torsion and non-metricity based analogues of NCT. One can shift between the theories by introducing a change in connection just as is done in the relativistic trinity. One helpful way of seeing this is to consider the general non-relativistic limit of \eqref{decomp}, which gives the difference tensor between the standard NCT connection $\hat{\Gamma}^a{ }_{bc}$ (which is the non-relativistic limit of the GR Levi-Civita connection \parencite{1976_Kunzle}) and a general affine connection $\tilde{\Gamma}^a{ }_{bc}$ \parencite[Eq.~32]{Wolf:2023rad}:
\begin{align}
    \hat{\Gamma}^a{ }_{bc}-\tilde{\Gamma}^a{ }_{bc}&=
        h^{s a}\left(\nabla_{(b} \hat{h}_{c)s} - \frac{1}{2} \nabla_s \hat{h}_{bc}\right) + \xi^a \nabla_{(b} t_{c)}  \nonumber \\
        &\qquad- \frac{1}{2} T^{\,a}{}_{bc} - h^{a s}T^{\,g}{}_{s (b} \hat{h}_{c)g} + 2h^{an}t_{(b}f_{c)n} + \mathcal{O}(\lambda),
\end{align}
where $\nabla_a t_b = Q_{ab}$ and $\nabla_a \hat{h}_{bc}= Q_{abc}$ refer to the temporal non-metricity and the non-metricity of the spatial projector (the other non-metric degrees of freedom are given by $\nabla_a h^{bc}=Q\indices{_a^b^c}$ and $\nabla_a \xi^b = Q\indices{_a^b}$ but do not explicitly appear in this version of the expression---see \textcite[Sect.~3.4]{Wolf:2023rad}), $T$ is the torsion, and $f_{ab}$ is a two-form (restricted to satisfy $f_{ab}=t_{[a}\nabla_{b]}\phi$ from the non-relativistic limit---see \textcite{Wolf:2023rad}). In order to obtain the fully generic version of TENC obtained in 
\textcite{Schwartz_2023}, one sets the non-metricities in the above formula to zero, thereby obtaining a contorsion tensor:
\begin{align}
    K\indices{^{a}_{bc}} =
    - \frac{1}{2} T^{\,a}{}_{bc} - h^{a s}T^{\,g}{}_{s (b} \hat{h}_{c)g} + 2h^{an}t_{(b}f_{c)n}.\label{eq:contorsion}
\end{align}
Similarly, in order to obtain the fully generic version of STENC in \textcite{Wolf:2023rad}, one sets all the torsions to zero to obtain the distortion tensor:
\begin{align}
    L\indices{^{a}_{bc}} =
    h^{s a}\left(\nabla_{(b} \hat{h}_{c)s} - \frac{1}{2} \nabla_s \hat{h}_{bc}\right) + \xi^a \nabla_{(b} t_{c)}  + 2h^{an}t_{(b}f_{c)n}.\label{eq:distortion}
\end{align}

TENC is then a theory given by models of the form $\langle M, t_a, h^{ab}, \overset{t}{\nabla}, \Phi \rangle$, where $\overset{t}{\nabla}$ is a flat, metric-compatible derivative operator with non-vanishing torsion. As discussed above, one can translate between NCT  and TENC is given by the following shift of connection $\overset{t}{\nabla} = (\overset{c}{\nabla}, K\indices{^{a}_{bc}} )$. It has been well-known since the work of \textcite{Trautman} that one can translation between NCT and NGT (in the form of the the famous `geometrisation' and `recovery' theorems---(see also \textcite[Ch.~4]{Malament}); however, it was much more recently shown by \textcite{ReadTeh} that standard Newtonian gravity can be understood as the `teleparallel equivalent' of NCT in much the same way that TEGR is the teleparallel equivalent of GR. In order to see this, one can `fix' the various torsional degrees of freedom. In particular, there are torsional degrees of freedom in temporal torsion, spatial torsion, and `mass' torsion (i.e.,\ the `mass gauge field' $m_a$ is obtained by gauging the Bargmann algebra \parencite{Andringa:2010it}). As discussed by \textcite[Sect.~4.1]{Schwartz_2023}, one can `fix' purely spatial torsion to vanish, while imposing some additional constraints on the temporal and mass torsions (i.e., $t_nT\indices{^n_b_c}=0$ and $T(M) = f_{ab}=t_{[a}\overset{t}{\nabla}_{b]}\phi $). After doing so, we formally recover the dynamics of NGT from TENC:
\begin{equation}\label{TENCeqn}
    \overset{t}{\nabla}_a \overset{t}{\nabla}\tensor{\vphantom{\nabla}}{^a} \phi = 4 \pi  \rho.
\end{equation}
In other words, TENC interprets Newtonian gravity as a result of a torsional connection.

Even more recently, STENC has been constructed by \textcite{Wolf:2023rad}. This theory is then given by models of the form $\langle M, t_a, h^{ab}, \overset{n}{\nabla}, \Phi \rangle$, where $\overset{n}{\nabla}$ is flat and torsion-free, but possesses non-vanishing non-metricity. Similarly, one can translate between NCT and STENC by introducing the difference tensor $\overset{n}{\nabla} = (\overset{c}{\nabla}, L\indices{^{a}_{bc}})$.
As with NCT and TENC above, one can obtain an equation formally equivalent to NGT, but with the gravitational field resulting from a connection that is flat and torsionless, but that encodes non-metricity. In STENC, there are non-metricity degrees of freedom in the non-metricities of the temporal metric $t$, spatial metric $h$, the spatial projector $\hat{h}$, and the velocity vector field $v$. As discussed by \textcite[Sect.~4.3]{Wolf:2023rad}, one can `fix' the non-metricities of the spatial projector and velocity vector field to vanish (i.e., $\overset{n}{\nabla}_a \hat{h}_{bc}= Q_{abc} = \overset{n}{\nabla}_a \xi^b = Q\indices{_a^b} = 0$), which as before recovers the familiar dynamics:
\begin{equation}\label{STENCeqn}
    \overset{n}{\nabla}_a \overset{n}{\nabla}\tensor{\vphantom{\nabla}}{^a} \phi = 4 \pi  \rho.
\end{equation}
As promised, this is formally equivalent to NGT, but results from a connection with non-metricity.

This `non-relativistic geometric trinity' mirrors the more familiar relativistic geometric trinity in that one can present three gravitational theories that are dynamically equivalent to familiar Newtonian gravitational theory, formulated in the geometric languages of curvature, torsion, and non-metricity.\footnote{To shore this up,  one would like to see geometrisation/recovery theorems \emph{à la} Trautman for all three nodes of the non-relativistic trinity. This turns out to be a delicate business, which we will address in a companion paper. The lack of explicit presentation of any such theorems in this article does not detract from the technical or conceptual points which we seek to make.} While in the relativistic case this is apparent at the level of the action, NCT, TENC, and STENC by contrast cannot be formulated using an action principle (for the reasons underlying this, see \textcite{ObersAction}), so we can only demonstrate their equivalence at the level of equations of motion.\footnote{Off-shell non-relativistic equivalence would require recourse to the `type II' versions of these theories, which we discuss in \S\ref{sec:conc}.} However, the non-relativistic trinity bears another important relationship to the relativistic trinity. All of the theories in the non-relativistic trinity can be obtained by taking an appropriate non-relativistic limit (typically in terms of a $1/c^2$ expansion in the style of \textcite{Schwartz_2023}). That is, NCT is the non-relativistic limit of GR \parencite{1976_Kunzle}, TENC is the non-relativistic limit of TEGR \parencite{Schwartz_2023}, and STENC is the  non-relativistic limit of STEGR \textcite{Wolf:2023rad}.
This completes all the legs of Figure \ref{fig:1}.

\section{Maxwell gravitation and the non-relativistic trinity}\label{sec:Maxwell}
As emphasised in the previous section, the three nodes of each of the relativistic and non-relativistic geometric trinities are all empirically equivalent theories, which nevertheless appear to disagree fundamentally on the geometrical structure which they attribute to the world. For instance, according to GR the gravitational behaviour of matter is to be understood as a manifestation of spacetime curvature, whereas according to TEGR and STEGR spacetime is necessarily flat. This means that the relativistic and non-relativistic theories present a case of strong underdetermination---distinct theories between which no possible evidence could be expected to decide.\footnote{\textcite{Knox2011-KNONTA} argues that there is in fact no strong underdetermination here; see \textcite{RuwardRead, WolfReadSanchioni} for responses, and \S\ref{sec:relcommoncore} for further discussion in this article.}

Faced with such cases of strong underdetermination, philosophers have suggested several approaches to dealing with the problem (on this see e.g.~\textcite{BLBJR}). Famously, one of these is the common core approach.
The common core approach advocates identifying the invariant kinematical structure of the theories, and then showing that this structure is sufficient to formulate a distinct, ontologically viable theory in its own right; one which, moreover, retains the empirical content of the original theories. Moving to this new interpretative framework alongside a judicious invocation of Occamist reasoning (on which see \textcite{Dasgupta}) then allows one to `break' the underdetermination by interpreting the theories in such a way that they completely agree on the structure they attribute to the world. The aim of this section is to show that in the case of the non-relativistic geometric trinity, such a common core theory exists, and it is a theory known as `Maxwell gravitation' (MG): a theory which has quite independently attracted philosophical interest (for reasons to do with (1) as presented in the introduction).

To do so, we begin by recalling some facts about Maxwellian spacetime. This is a structure $\langle M, t_a, h^{ab}, \circlearrowright\rangle$, where $t_a$, $h^{ab}$ are orthogonal temporal and spatial metrics as introduced in the previous section, and $\circlearrowright$ is a standard of rotation compatible with $t_a$ and $h^{ab}$. This was introduced originally by \textcite{Weatherall7}: if $t_a$, $h^{ab}$ are compatible temporal and spatial metrics on $M$, then a standard of rotation $\circlearrowright$ compatible with $t_a$ and $h^{ab}$ is a map from smooth vector fields $\xi^a$ on $M$ to smooth, antisymmetric rank-$(2,0)$ tensor fields $\circlearrowright^b\!\xi^a$ on $M$, such that
\begin{enumerate}
    \item $\circlearrowright$ commutes with addition of smooth vector fields;
    \item Given any smooth vector field $\xi^a$ and smooth scalar field $\alpha$, $\circlearrowright^a\!(\alpha\xi^b)=\alpha\circlearrowright^a\!\xi^b+\xi^{[b}d^{a]}\alpha$;
    \item $\circlearrowright$ commutes with index substitution;
    \item Given any smooth vector field $\xi^a$, if $d_a(\xi^nt_n)=0$ then $\circlearrowright^a\!\xi^b$ is spacelike in both indices; and
    \item Given any smooth spacelike vector field $\sigma^a$, $\circlearrowright^a\!\sigma^b=D^{[a}\sigma^{b]}$,
\end{enumerate}
where $D$ is the unique Levi-Civita connection induced by $h^{ab}$ on each spacelike hypersurface. We will say that a connection and a standard of rotation are \textit{compatible} iff $\circlearrowright^a\eta^b=\nabla^{[a}\eta^{b]}$ for all vector fields $\eta^a$ on $M$. It follows that any metric compatible, torsion-free connection $\nabla$ on $M$ determines a unique compatible standard of rotation---namely, the map $\circlearrowright:\eta^a\rightarrow\nabla^{[a}\eta^{b]}$  \parencite{Weatherall7}.


However, in light of the discussion of the previous section, this invites a natural further question: are there non-metric or torsionful connections which are \textit{also} associated with metric compatible standards of rotation in the above sense? It turns out that the answer to this question is `yes', and a characterisation of such connections is given by the following two propositions:
\begin{prop}\label{prop:nonmetricrotation}
    Let $\langle M, t_a, h^{ab}, \nabla\rangle$ be a non-relativistic spacetime, where $t_a$ and $h^{ab}$ are compatible, $t_a$ is closed,\footnote{The temporal metric being closed is a standard assumption in much of the non-relativistic gravity literature. This is due to a technical fact that it is only the zeroth order in the expansion of the Levi-Civita connection that transforms as a connection. One can then \textit{impose} that $dt=0$ so that the minus-first order vanishes; the zeroth order of the expansion then becomes the leading order and can serve as a proper connection for the Galilean theory (see e.g.~\textcite{kun1, VandenBleeken:2017rij}). While this is not the only option, the choice of $dt=0$ is the most straightforward way of taking the non-relativistic limit of GR. Furthermore, this choice preserves a notion of absolute time, which will be the case with theories of non-relativistic gravity that can be understood to geometrise standard Newtonian gravity, which are precisely the kinds of theories with which the present work is concerned.} and where $\nabla_at_b=Q_{ab}$ and $\nabla_ah^{bc}=Q\indices{_a^b^c}$. Then the map $\circlearrowright:\eta^a\rightarrow\nabla^{[a}\eta^{b]}$ is a standard of rotation compatible with $t_a$ and $h^{ab}$ iff $h^{an}Q_{nb}=h\indices{^{n[a}}Q\indices{_n^{b]}^c}=0$.
\end{prop}
\begin{proof}
    First, suppose that $h^{an}Q_{nb}=h^{n[a}Q\indices{_n^{b]}^c}=0$. That $\circlearrowright$ satisfies conditions (1)--(3) is immediate from properties of derivative operators. For condition (4), note that if $d_a(\eta^nt_n)=0$ we have
    \begin{equation}
        0=\nabla_a(\eta^nt_n)=t_n\nabla_a\eta^n+\eta^nQ_{na}, \label{1}
    \end{equation}
    so that
    \begin{align*}
        t_n(h^{m[n}\nabla_m\eta^{a]})&=-\frac{1}{2}h^{ma}t_n\nabla_m\eta^n\\
        &=\frac{1}{2}h^{ma}\eta^nQ_{mn}\\
        &=0\\
        &=t_n(h^{m[a}\nabla_m\eta^{n]})
\end{align*}
    and $\circlearrowright^a\!\eta^b$ is spacelike in both indices. Finally, consider condition (5). We know that $\nabla^{[a}h^{b]c}=h^{n[a}Q\indices{_n^{b]}^c}=0$. So let $\xi^a$ be a unit timelike vector field on $M$, $\hat{h}_{ab}$ the spatial metric relative to $\xi^a$,\footnote{That is, the unique symmetric tensor field on $M$ such that $\hat{h}_{an}\xi^n=0$ and $h^{an}\hat{h}_{nb}={\delta^a}_b-t_b\xi^a$.} and $D$ the unique spatial derivative operator such that $D_ah^{bc}=0$. Then for any spacelike vector field $\sigma^a$ on $M$,
    \begin{equation}
        h^{n[a}D_n\sigma^{b]}=h^{n[a}\hat{h}_{nm}\tensor{\hat{h}}{^{b]}_r}\nabla^m\sigma^r=h^{n[a}\nabla_n\sigma^{b]}, \label{2}
    \end{equation}
    where the first equality follows from the proof of \parencite[Proposition 2]{Weatherall7}, and we have used the fact that since $h^{an}Q_{nb}=0$, $\nabla^a(t_n\eta^n)=t_n\nabla^a\eta^n$ for any smooth vector field $\eta^a$ on $M$.\\
    Conversely, suppose that the map $\circlearrowright:\eta^a\rightarrow\nabla^{[a}\eta^{b]}$ is compatible with $t_a$ and $h^{ab}$. Then we must have for all smooth vector fields $\eta^a$ on $M$ that if $d_a(\eta^nt_n)=0$, $\circlearrowright^a\!\eta^b$ is spacelike in both indices. If $d_a(\eta^nt_n)=0$ then \eqref{1} holds with respect to $\eta^a$ so that $t_n(h^{m[n}\nabla_m\eta^{a]})=1/2h^{ma}Q_{mn}\eta^n=0$. But this can only be the case for arbitrary $\eta^a$ if $h^{an}Q_{nm}=0$. Moreover, given condition (5) we must then also have that the equality \eqref{2} holds with respect to $\nabla$. Now, we know that the action of $D$ on spacelike vector fields $\sigma^a$ is defined as follows: $h^{n[a}D_n\sigma^{b]}=h^{n[a}\hat{h}_{nm}\tensor{\hat{h}}{^{b]}_r}\nabla'^m\sigma^r$ for any $\nabla'$ such that $\nabla'^ah^{bc}=0$.\footnote{This follows from the proof of proposition 2 of \textcite{Weatherall7}.} In particular, this is the case for $\nabla'=(\nabla, -1/2\hat{h}_{bn}\hat{h}_{cm}(h^{ar}Q\indices{_r^n^m}-h^{nr}Q\indices{_r^a^m}-h^{mr}Q\indices{_r^n^a}))$. 
    Thus 
    \begin{align*}
        0&=h^{n[a}\hat{h}_{nm}\tensor{\hat{h}}{^{b]}_r}h^{ms}\hat{h}_{st}\hat{h}_{vu}(h^{rw}Q\indices{_w^t^u}-h^{tw}Q\indices{_w^r^u}-h^{uw}Q\indices{_w^t^r})\sigma^v\\
        &=\hat{h}\indices{^{[a}_t}\tensor{\hat{h}}{^{b]}_r}\hat{h}_{vu}(h^{rw}Q\indices{_w^t^u}-h^{tw}Q\indices{_w^r^u}-h^{uw}Q\indices{_w^t^r})\sigma^v\\
        &=2\delta\indices{^{[a}_t}\delta\indices{^{b]}_r}\hat{h}_{vu}h^{w[r}Q\indices{_w^{t]}^u}\sigma^v\\
        &=h^{w[b}Q\indices{_w^{a]}^u}\sigma_u
    \end{align*}
    for any covector $\sigma_a$, where we have used that the metrics are orthogonal (so that $t_nQ\indices{^a^n^b}$ vanishes if $Q\indices{^a_b}$ does) and that $Q\indices{_a^b^c}$ is symmetric in the upper two indices. But this can only be the case for arbitrary $\sigma_a$ if $h^{n[a}Q\indices{_n^{b]}^c}=0$.
\end{proof}
\begin{prop}\label{prop:torsionrotation}
    Let $\langle M, t_a, h^{ab}, \nabla\rangle$ be a non-relativistic spacetime, where $t_a$ and $h^{ab}$ are compatible, $t_a$ is closed, and $\nabla$ is compatible with the metrics but possibly torsionful. Then the map $\circlearrowright:\eta^a\rightarrow\nabla^{[a}\eta^{b]}$ is a standard of rotation compatible with $t_a$ and $h^{ab}$ iff $T^{abc}=0$.  
\end{prop}
\begin{proof}
    First, note that since $\nabla$ is compatible with $t_a$ and $t_a$ is closed, $t_n{T^n}_{ab}=0$. That the map $\circlearrowright:\eta^a\rightarrow\nabla^{[a}\eta^{b]}$ satisfies conditions (1)-(3) is again immediate from properties of derivative operators. (4) follows from the fact that $\nabla$ is compatible with $t_a$, using that $d_a\alpha=\nabla_a\alpha$ for any 0-form field $\alpha$. Finally, consider (5). Let $\xi^a$ be a unit timelike vector field on $M$, and $\hat{h}_{ab}$ the spatial metric relative to $\xi^a$. We know that the action of $D$ on spacelike vector fields is defined as follows: $D_a\sigma^b=\hat{h}_{an}\tensor{\hat{h}}{^b_m}\nabla'^n\sigma^m$, where $\nabla'$ is an arbitrary torsion-free derivative operator such that $\nabla'^ah^{bc}=0$. Moreover, we know (equation \eqref{eq:contorsion}) that $\nabla=(\nabla', {K^a}_{bc})$ for some such $\nabla'$. Hence
    \begin{align*}
        D^{[a}\sigma^{b]}&=\tensor{\hat{h}}{^{[a}_n}\tensor{\hat{h}}{^{b]}_m}\nabla'^n\sigma^m\\
        &=\tensor{\hat{h}}{^{[a}_n}\tensor{\hat{h}}{^{b]}_m}(\nabla^n\sigma^m-h^{nr}{K^m}_{rs}\sigma^s)\\
        &=\nabla^{[a}\sigma^{b]}-{K^{[ab]}}_n\sigma^n\\
        &=\nabla^{[a}\sigma^{b]}-\frac{1}{2}T^{nab}\sigma_n
    \end{align*}
    for some covector $\sigma_n$, where we have used the fact that $\sigma^a$ is spacelike and $t_n{T^n}_{ab}=0$. Thus if $T^{abc}=0$, then (5) is satisfied. Conversely, if the map $\circlearrowright:\eta^a\rightarrow\nabla^{[a}\eta^{b]}$ satisfies (5), it follows that $T^{abc}=0$. 
\end{proof}
Propositions \ref{prop:nonmetricrotation} and \ref{prop:torsionrotation} are tantalising, because they show that non-relativistic affine connections which exhibit either torsion or non-metricity may---under certain conditions---be associated with a compatible standard of rotation, just as with curvature based connections. This raises the prospect that Maxwellian spacetime might be the invariant kinematic structure of the non-relativistic geometric trinity. We isolate the sense in which this is so in the following proposition:
\begin{prop} \label{prop:GTMaxwellian}
    Let $\langle M, t_a, h^{ab}\rangle$ be a non-relativistic spacetime, with $t_a$ and $h^{ab}$ defined as above, and consider three connections: a curvature based connection $\overset{c}{\nabla}$, a non-metricity based connection $\overset{n}{\nabla}$ with $\overset{n}{\nabla}\tensor{\vphantom{\nabla}}{_a}t_b=Q_{ab}$ and $\overset{n}{\nabla}\tensor{\vphantom{\nabla}}{_a}h^{bc}=Q\indices{_a^b^c}$, and a torsion based connection $\overset{t}{\nabla}$. Let $\overset{c}{\nabla}=(\overset{n}{\nabla}, {L^a}_{bc})=(\overset{t}{\nabla}, {K^a}_{bc})$, where ${L^a}_{bc}$ and ${K^a}_{bc}$ are given in \eqref{eq:distortion} and \eqref{eq:contorsion} respectively. Suppose that $h^{an}Q_{nb}=h^{n[a}Q\indices{_n^{b]}^c}=0$ and $T^{abc}=0$. Let $\xi^a$ be the unit timelike vector field with respect to which ${L^a}_{bc}$ and ${K^a}_{bc}$ are defined. Then if $\xi^a$ is twist-free with respect to $\overset{c}{\nabla}$, $\xi^a$ is twist-free with respect to $\overset{n}{\nabla}$ and $\overset{t}{\nabla}$ iff $f_{ab}=t_{[a}\phi_{b]}$ for some covector $\phi_a$.
\end{prop}
\begin{proof}
    First, we note that the expression \eqref{eq:distortion} for ${L^a}_{bc}$ can equivalently be rewritten as (see \S\ref{app:distorsion} for details):
    \begin{align*}
        {L^a}_{bc}=-\frac{1}{2}\hat{h}_{bn}\hat{h}_{cm}&(h^{ar}Q\indices{_r^n^m}-h^{nr}Q\indices{_r^a^m}-h^{mr}Q\indices{_r^n^a})+\xi^nt_{(b}\hat{h}_{c)m}Q\indices{_n^m^a}\\
        &-\xi^a\xi^nt_{(b}Q_{c)n}+2t_{(b}f_{c)n}h^{na}+2t_{(b}g_{c)n}h^{na},
    \end{align*}
    where $g_{ab}=\hat{h}_{n[b}\overset{n}{\nabla}\vphantom{\nabla}_{a]}\xi^n$. Thus
    \begin{align*}
        \tensor{L}{^{[ab]}_c}&=-\frac{1}{2}(\hat{h}_{nc}h^{r[a}Q\indices{_r^{b]}^n}-\xi^nt_c\hat{h}\indices{^{[b}_m}Q\indices{_n^{a]}^m}+\xi^n\xi^{[a}t_cQ\indices{^{b]}_n}-2t_cf^{[ab]}-2t_cg^{[ab]})\\
        &=-\frac{1}{2}(\xi^n\xi^{[b}t_ct_mQ\indices{_n^{a]m}}-\xi^nt_cQ\indices{_n^{[ab]}}-2t_cf^{ab}+2t_c\hat{h}\indices{_n^{[a}}\overset{n}{\nabla}\vphantom{\nabla}^{b]}\xi^n)\\
        &=-t_c(\overset{n}{\nabla}\vphantom{\nabla}^{[a}\xi^{b]}-\xi^n\xi^{[a}Q\indices{^{b]}_n}-f^{ab})\\
        &=-t_c(\overset{n}{\nabla}\vphantom{\nabla}^{[a}\xi^{b]}-f^{ab}),
    \end{align*}
    where we have used in the second equality that $h^{an}Q_{nb}=h^{n[a}Q\indices{_n^{b]}^c}=0$ and that $f_{ab}$ is antisymmetric, in the third equality that $Q\indices{_a^{[bc]}}=0$ and that $h^{an}Q_{nb}=0$ implies $t_nh^{am}Q\indices{_m^b^n}=0$, and in the fourth equality again that $h^{an}Q_{nb}=0$. 
    
    But if $\overset{c}{\nabla}\vphantom{\nabla}^{[a}\xi^{b]}=0$, then $\overset{n}{\nabla}\vphantom{\nabla}^{[a}\xi^{b]}=\tensor{L}{^{[ab]}_n}\xi^n=-(\overset{n}{\nabla}\vphantom{\nabla}^{[a}\xi^{b]}-f^{ab})$ and hence $\overset{n}{\nabla}\vphantom{\nabla}^{[a}\xi^{b]}=\frac{1}{2}f^{ab}$. It follows that $\overset{n}{\nabla}\vphantom{\nabla}^{[a}\xi^{b]}=0$ iff $f^{ab}=0$. But this is equivalent to the requirement that $f_{ab}=t_{[a}\phi_{b]}$ for some covector $\phi_a$.

    Next, note that
    \begin{align*}
        \tensor{K}{^{[ab]}_c}&=\frac{1}{2}({T_c}^{ab}+2t_cf^{[ab]})\\
        &=\frac{1}{2}(\hat{h}_{cm}T^{mab}+2t_cf^{ab})\\
        &=t_cf^{ab},
    \end{align*}
    where we have used again that $f_{ab}$ is antisymmetric in the second equality and that $T^{abc}=0$ in the third. So if $\overset{c}{\nabla}\tensor{\vphantom{\nabla}}{^{[a}}\xi^{b]}=0$, $\overset{t}{\nabla}\tensor{\vphantom{\nabla}}{^{[a}}\xi^{b]}=\tensor{K}{^{[ab]}_n}\xi^n=f^{ab}$, which completes the proof by the same reasoning as above.
\end{proof}
\begin{corollary}\label{corr:GTMaxwellian}
    Let $\langle M, t_a, h^{ab}\rangle$ be a non-relativistic spacetime, with $t_a$ and $h^{ab}$ defined as above, and consider three connections: a curvature based connection $\overset{c}{\nabla}$, a non-metricity based connection $\overset{n}{\nabla}$ with $\overset{n}{\nabla}\tensor{\vphantom{\nabla}}{_a}t_b=Q_{ab}$ and $\overset{n}{\nabla}\tensor{\vphantom{\nabla}}{_a}h^{bc}=Q\indices{_a^b^c}$, and a torsion based connection $\overset{t}{\nabla}$. Let $\overset{c}{\nabla}=(\overset{n}{\nabla}, {L^a}_{bc})=(\overset{t}{\nabla}, {K^a}_{bc})$, where ${L^a}_{bc}$ and ${K^a}_{bc}$ are given in \eqref{eq:distortion} and \eqref{eq:contorsion} respectively. Suppose that $h^{an}Q_{nb}=h^{n[a}Q\indices{_n^{b]}^c}=0$, $T^{abc}=0$, and $f_{ab}=t_{[a}\phi_{b]}$ for some covector $\phi_a$. Let $\xi^a$ be the unit timelike vector field with respect to which ${L^a}_{bc}$ and ${K^a}_{bc}$ are defined, and suppose that $\overset{c}{\nabla}\tensor{\vphantom{\nabla}}{^{[a}}\xi^{b]}=\pounds_\xi h^{ab}=0$. Further suppose that $h^{ab}$ is flat.\footnote{In the sense that for all spacelike hypersurfaces $S$, $D$ commutes on spacelike vector fields, where $D$ is the unique spatial derivative operator such that $D_ah^{bc}=0$.} Then $\overset{c}{\circlearrowright}$, $\overset{n}{\circlearrowright}$, $\overset{t}{\circlearrowright}$ are standards of rotation compatible with the metrics and $\overset{n}{\circlearrowright}=\overset{c}{\circlearrowright}=\overset{t}{\circlearrowright}$. 
\end{corollary}
\begin{proof}
    This follows immediately from propositions \ref{prop:nonmetricrotation}, \ref{prop:torsionrotation}, \ref{prop:GTMaxwellian} and proposition 1 of \parencite{Weatherall7}, using that flat, compatible, torsion-free connections always determine the same standard of rotation. 
\end{proof}
Our claim that Maxwellian spacetime is the invariant kinematic structure of the non-relativistic geometric trinity then rests on the status of four conditions: (i) that $h^{ab}$ is flat, (ii) that $f_{ab}=t_{[a}\phi_{b]}$, (iii) that the unit timelike vector field $\xi^a$ with respect to which the contorsion and distortion tensors are defined satisfies $\overset{c}{\nabla}\tensor{\vphantom{\nabla}}{^{[a}}\xi^{b]}=\pounds_\xi h^{ab}=0$, and (iv) that $h^{an}Q_{nb}=h^{n[a}Q\indices{_n^{b]}^c}=0$, $T^{abc}=0$. We will now remark on each of these conditions in turn. 

Beginning with (i), this is entailed by the geometrised Poisson equation of NCT \parencite[proposition 4.1.5]{Malament}, and so holds automatically in all three nodes of the non-relativistic geometric trinity. For (ii), we have already noted in \ref{sec:nonrelGT} that the tensor $f_{ab}$ is restricted to take this form from the non-relativistic limit; thus, restricting to the versions of TENC and STENC that can be understood as the non-relativistic limits of TEGR and STEGR (respectively) is sufficient to ensure that this condition is satisfied. 

For (iii), the fact that there exists a unit timelike vector field such that $\overset{c}{\nabla}\tensor{\vphantom{\nabla}}{^{[a}}\xi^{b]}=\pounds_\xi h^{ab}=0$ follows from the second Trautman condition \eqref{trautman2} of NCT, so (iii) can always consistently be imposed. As justification for (iii), note that the unit timelike vector field with respect to which the contorsion and distorsion tensors are defined receives (roughly) an interpretation as an `inertial' observer, which experiences a four-acceleration $\xi^n\overset{c}{\nabla}\vphantom{\nabla}_n\xi^a-\hat{h}\indices{^a_m}\xi^n\overset{n}{\nabla}\vphantom{\nabla}_n\xi^m=\xi^n\overset{c}{\nabla}\vphantom{\nabla}_n\xi^a-\xi^n\overset{t}{\nabla}\vphantom{\nabla}_n\xi^a=2\xi^nf_{nm}h^{ma}=\phi^a$ due to gravitational interactions in NCT. So insofar as (iii) is supposed to define what we mean by an inertial observer in NCT (see, e.g.~\textcite{Wolf:2023rad, TehNG, Schwartz_2023}), this restriction is well-motivated. 

Finally, consider the condition (iv). One can show (see appendix \ref{app:spatial}) that necessary and sufficient for $\overset{t}{\nabla}$ and $\overset{n}{\nabla}$ to agree with the unique compatible torsion-free spatial derivative operator $D$ when projected onto each spacelike hypersurface $S$ are the conditions $h^{an}Q_{nb}=h^{na}Q\indices{_n^b^c}=0$, $T^{abc}=0$. In turn, this agreement with $D$ is necessary for the recovery of the Poisson equation in TENC or STENC. So insofar as one understands TENC and STENC as theories in which one can derive the Poisson equation, (iv) must be included as an additional assumption. 

To summarise the results of this section so far, then: we've isolated a Maxwellian spacetime structure as the common core of the non-relativistic geometric trinity. The next point to note is that this structure is sufficient to formulate the dynamics of Newtonian gravity. This was first shown by \textcite{Dewar}, and recently given an `intrinsic' formulation by \textcite{March, Chen}; the resulting theory---`Maxwell gravitation'---has models of the form $\langle M, t_a, h^{ab}, \circlearrowright, \Phi\rangle$. 
Together with propositions \ref{prop:nonmetricrotation}, \ref{prop:torsionrotation}, and \ref{prop:GTMaxwellian}, and corollary \ref{corr:GTMaxwellian} this substantiates our earlier claim that MG constitutes the dynamical common core of the non-relativistic geometric trinity.
It also paves the way for an interpretation of Newtonian gravity on which the structure it attributes to the world is strictly \textit{less} than that of a connection.

This takes us to the connection with Corollary VI and the Trautman symmetry, to which we alluded in Section \ref{sec:intro}. As articulated by \textcite{Jacobs2023}, the `dynamic shift' symmetry of potential-based Newtonian gravity \textit{\`a la} Corollary VI and the Trautman symmetry in which the connection and gravitational potential are altered simultaneously can be understood as being two sides of the same coin: both consequences of the invariance of the Newtonian dynamics under the Maxwell group. But Maxwell transformations produce a linear, time-dependent acceleration of the matter content of the original solution. \textit{Prima facie}, one might think that purging the theory of this symmetry would involve excising the structure needed to make sense of such linear accelerations---to wit, a connection---leaving only the standard of rotation.

In that sense, that MG should be the dynamical common core of the non-relativistic geometric trinity was already suggested by the dynamical symmetries of Newtonian gravity. On the one hand, the irrotational degrees of freedom of the connection were already known to be superfluous to the internal dynamics of the matter distribution. On the other hand, agreement on the rotation standard is necessary for empirical equivalence to standard Newtonian gravity.

What, then, to make of the fact that one can also use Trautman symmetry to motivate the move to NCT? One way to think about this is that while we are always free to \textit{define} a connection from the standard of rotation and matter distribution by coupling the degrees of freedom of the connection to the matter distribution, there is necessarily a certain amount of slack in how this connection is constructed. This is because the projective degrees of freedom of an affine connection far outstrip the degrees of freedom of the matter distribution. Taking up this slack in different places allows us to express Newtonian gravity as a theory of curvature, or torsion, or non-metricity---in some cases, we can even specify the connection uniquely! That we can specify the curvature-based connection uniquely under certain weak conditions on the mass density field is what ensures that $\overset{c}{\nabla}$, as well as the rotation standard, is also an invariant of Trautman gauge symmetry. But the fact remains \textit{viz-\`a-viz} Corollary VI that the full structure of an affine connection is not needed to support the dynamics, and so, if one introduces such a connection, one has to reckon with the fact that---again, necessarily---there are multiple distinct ways of doing so. 

\section{GR as the relativistic common core}\label{sec:relcommoncore}

Recall that the projective structure of a spacetime theory identifies a certain subset of worldlines as the (unparameterised) geodesics; the conformal structure of a given relativistic spacetime theory specifies a lightcone at every spacetime point (see e.g.~\textcite{MatveevScholz}). Famously---a result going back to \textcite{Weyl1921}---a Lorentzian spacetime is fixed by its associated projective and conformal structure: see \textcite[ch.~2]{Malament}; the corresponding existence result was proved by \textcite{EPS}, and is discussed further by \textcite{LinnemannReadEPS}. \textcite{WolfReadSanchioni} identify that one can move between nodes of the relativistic geometric trinity by modifying projective structure while leaving conformal structure unchanged (for all three theories leave lightcone structure unmodified); in a similar manner (one ultimately irrelevant to our purposes here, but perhaps nevertheless worth pointing out) one can move to non-relativistic theories by `widening the lightcone', thereby changing conformal structure (this constitutes a geometrical way of thinking about taking the non-relativistic limit: see \textcite{Malament1986}).

The three nodes of the relativistic geometric trinity then all agree on conformal structure. But the reason for this is that they all agree on metric structure (see \S\ref{sec:relGT}). This means that the invariant kinematical structure of the relativistic geometric trinity is $\langle M, g_{ab}, \Phi\rangle$. Is this structure sufficient to formulate the dynamics of the relativistic trinity---i.e.~a dynamical common core?

We claim that the answer to this is `yes', on at least one plausible way of understanding what the `structure' associated to a theory is. For this, recall that the Levi-Civita connection of GR is \textit{definable} from the metric. So if one understands the `structure' of a theory to include (a) all those structures to which the theory is explicitly committed---i.e.~which show up between the angle brackets of its models---and (b) any structures which are definable therefrom, then the relativistic geometric trinity does indeed have a dynamical common core, which is just GR---\emph{unlike} the non-relativistic geometric trinity (the dynamical common core of which is MG), the common core is not a distinct theory.

Several points are in order here. First one might question whether the above claim about the structural commitments of theories is correct. Second, given that $\langle M, g_{ab}, \Phi\rangle$ is the invariant kinematical structure of the relativistic geometric trinity, one might quite reasonably ask whether it is the case that the structure of MG can be, in some sense, seen to arise as the non-relativistic limit of this. (This is especially pressing given that in \S\ref{sec:nonrelGT} we identified NCT, rather than MG, as the non-relativistic limit of GR.) Third, if the dynamical common core of the relativistic trinity is GR, then one might wonder about the connection to Knox's apparently similar claim that in TEGR, ``the old entities from GR appear to be waiting in the wings'' \parencite[p.~273]{Knox2011-KNONTA}. We'll now address each of these in turn.

Beginning with the first, we claim that the above characterisation of `structure' (which follows the lead of \textcite{Barrett2017-BARWDS-3}) is indeed a natural and plausible one, and one which, moreover, is assumed throughout philosophical discourse on the structure of theories. To expand on what we mean here, consider the following representative statements:
\begin{itemize}
    \item[(i)] A cotangent bundle $T^*M$ comes canonically equipped with a symplectic form, so a cotangent bundle has the structure of a symplectic manifold.
    \item[(ii)] From the electromagnetic $1$-form $A_a$ on $\langle M, \eta_{ab}\rangle$ we can define the Faraday tensor $F_{ab}$ but not \textit{vice versa}, therefore this 1-form has more structure than the Faraday tensor.
    \item[(iii)] Full Newtonian spacetime $\langle M, t_a, h^{ab}, \xi^a\rangle$ has more structure than Galilean spacetime $\langle M, t_a, h^{ab}, \nabla\rangle$; it has all the structure that Galilean spacetime does, but also has a standard of rest.
\end{itemize}
These statements are intuitively compelling, and moreover, strike us as \textit{true}. But unless our claim about what one should take the structure of a theory to include is correct, it is difficult to make sense of this. E.g.,\ on (i), nowhere in writing down `$T^*M$'
did we explicitly write down a symplectic form $\omega$ on $T^*M$! On the other hand, our claim about the structure of theories offers an elegant explanation of \textit{why} we should think these statements are true. E.g.,\ on (iii), the fact that full Newtonian spacetime has all the structure that Galilean spacetime has (despite the connection not showing up explicitly in its models) can be seen by noting that the Galilean connection is definable from the metrics and standard of rest (see \textcite[Proposition 4.2.4]{Malament}). 

Turning now to the second, consider the non-relativistic limit of $\langle M, g_{ab}, \Phi\rangle$. In the non-relativistic limit, the metric degenerates (see \textcite{Malament1986}), leaving us with $\langle M, t_a, h^{ab}, \Phi\rangle$. This structure is certainly not sufficient to formulate the dynamics of Newtonian gravitation theory---hence it is standard to take the non-relativistic limit of the formulation of GR presented in \S\ref{sec:relGT} which explicitly includes the connection, which gives us NCT. However, once we have taken this limit---which gives us the geometrised Poisson equation and \eqref{trautman1}---we still need to impose \eqref{trautman2} for full empirical equivalence with Newtonian gravity. What happens if we also impose \eqref{trautman2} on the spacetime $\langle M, t_a, h^{ab}, \Phi\rangle$?

The basic point here is that for it even to make sense to talk of imposing \eqref{trautman2}, we still need to introduce some extra structure to $\langle M, t_a, h^{ab}, \Phi\rangle$. But since \eqref{trautman2} is equivalent to the condition that (a) $h^{ab}$ is flat, and (b) there exists a unit timelike vector field on $M$ which is twist-free and rigid \parencite[Proposition 4.2.4]{Malament}, this structure is strictly \textit{less} than that of a connection---we only need the standard of rotation. So by taking the non-relativistic limit of $\langle M, g_{ab}, \Phi\rangle$, and then imposing the most minimal structure needed to make sense of the conditions for empirical equivalence to Newtonian gravity, one arrives naturally at the structure of Maxwellian spacetime. It is in this sense that the invariant kinematical structure of non-relativistic trinity---i.e.~Maxwellian spacetime---can be seen as the non-relativistic limit of the invariant kinematical structure of the relativistic trinity $\langle M, g_{ab}, \Phi\rangle$.

Finally, we move to our third point---\emph{viz.},~the connection with \textcite{Knox2011-KNONTA}. On this, it is worth emphasising that Knox's aims 
are somewhat different to ours---Knox is primarily concerned with whether GR and TEGR are in some sense \textit{equivalent} theories, rather than with identifying their common mathematical structure. Nevertheless, there is a close relationship between Knox's argument and our own. For example, whereas Knox takes the fact that the metric and Levi-Civita connection are still definable in TEGR (and moreover, that matter still couples to the Levi-Civita connection) to suggest that TEGR is a mere notational variant of GR, we have argued that this means that GR can be seen as the dynamical common core of the relativistic trinity---in which case, the dynamics of GR are indeed somehow `contained' in the dynamics of TEGR, which is very much in the spirit of Knox's point. Yet, \emph{contra} Knox, and as argued by \textcite{RuwardRead} and  \textcite{WolfReadSanchioni}, there are still important reasons to regard GR, TEGR, and STEGR as distinct theories and not merely notational variants of each other---the `surplus' structure that is present in TEGR and STEGR is seen by many as important for theoretical and physical reasons that lead them to be actively pursued as viable gravitational theories.

\section{Conclusions}\label{sec:conc}

In this article, we've shown that Maxwell gravitation constitutes the mathematical common core of the recently-discovered non-relativistic geometric trinity of gravity (first presented by \textcite{Wolf:2023rad}); we've also explained why the common core of the relativistic geometric trinity is GR, but this common core interpretation is less compelling than in the case of the non-relativistic trinity. In undertaking this work, we take ourselves to have made good on the exhortation of \textcite{Lehmkuhl3} to explore and chart the `space of spacetime theories'---at least with respect to this small (albeit philosophically important!)~corner of the landscape.


There are many future prospects; here we mention just two. First: one might be interested in whether (a) there exists an extended non-relativistic geometric trinity for the off-shell Newtonian limit presented by \textcite{Obers} (this question was also raised by \textcite{Wolf:2023rad}), and (b) if so, whether there exists a dynamical common core to this non-relativistic trinity. And second: one might wonder whether (a) there exists an \emph{ultra}-relativistic geometric trinity obtained by taking the ultra-relativistic (i.e.~roughly speaking, $c \rightarrow 0$) limit of the relativistic geometric trinity, and (b) again whether there exists a dynamical common core to that trinity also.

 \section*{Acknowledgements}

 We are grateful to Quentin Vigneron, to Jim Weatherall, and to the anonymous referees, for helpful feedback. E.M.~acknowledges support from Balliol College, Oxford, and the Faculty of Philosophy, University of Oxford. W.W.~acknowledges support from the Center for the History and Philosophy of Physics at St.~Cross College, Oxford and the British Society for the Philosophy of Science. J.R.~acknowledges the support of the Leverhulme Trust. 

\appendix

\section{Alternative expressions for the non-relativistic distorsion}\label{app:distorsion}
In this appendix, we show that the non-relativistic distorsion tensor can be written in a couple of different ways. First, note that
\begin{align*}
    \frac{1}{2}\hat{h}_{bn}\hat{h}_{cm}&(h^{ar}Q\indices{_r^n^m}-h^{nr}Q\indices{_r^a^m}-h^{mr}Q\indices{_r^n^a})\\
    &=\frac{1}{2}(\hat{h}_{bn}\hat{h}_{cm}h^{ar}Q\indices{_r^n^m}-(\delta\indices{_b^r}-t_b\xi^r)\hat{h}_{cm}Q\indices{_r^a^m}-(\delta\indices{_c^r}-t_c\xi^r)\hat{h}_{bn}Q\indices{_r^n^a})\\
    &=\frac{1}{2}(\hat{h}_{bn}\hat{h}_{cm}h^{ar}Q\indices{_r^n^m}-2\hat{h}_{n(b}Q\indices{_{c)}^n^a}+2\xi^rt_{(b}\hat{h}_{c)m}Q\indices{_r^m^a}).
\end{align*}
But
\begin{align*}
   L\indices{^{a}_{bc}}&=
    h^{s a}\left(\nabla_{(b} \hat{h}_{c)s} - \frac{1}{2} \nabla_s \hat{h}_{bc}\right) + \xi^a \nabla_{(b} t_{c)}  + 2h^{an}t_{(b}f_{c)n}\\
    &=-\frac{1}{2}h^{an}\hat{h}_{bm}\hat{h}_{cr}Q\indices{_n^m^r}+\hat{h}_{n(b}Q\indices{_{c)}^n^a}-\xi^a\xi^nt_{(b}Q_{c)n}+2h^{an}t_{(b}f_{c)n}+ h^{an}t_{(b}g_{c)n}
\end{align*}
(see Appendix B of \textcite{Wolf:2023rad} for derivation), so that
\begin{align*}
    {L^a}_{bc}=-\frac{1}{2}\hat{h}_{bn}\hat{h}_{cm}&(h^{ar}Q\indices{_r^n^m}-h^{nr}Q\indices{_r^a^m}-h^{mr}Q\indices{_r^n^a})+\xi^nt_{(b}\hat{h}_{c)m}Q\indices{_n^m^a}\\
    &-\xi^a\xi^nt_{(b}Q_{c)n}+2t_{(b}f_{c)n}h^{na}+2t_{(b}g_{c)n}h^{na}.
\end{align*}

\section{Restriction of torsionful and non-metric connections to spacelike hypersurfaces}\label{app:spatial}
In this appendix, we prove two propositions regarding the restriction of torsionful and non-metric connections to spacelike hypersurfaces. The first is this:
\begin{prop}
    Let $\nabla$ be a torsionful connection on $\langle M, t_a, h^{ab}\rangle$, and let $D$ denote the induced spatial derivative operator on each spacelike hypersurface $S$. Then $D$ is torsion-free iff $T^{abc}=0$.  
\end{prop}
\begin{proof}
    Let $\alpha$ be an arbitrary scalar field. We have that
    \begin{align*} 
2D_{[a}D_{b]}\alpha&=2\hat{h}\indices{^n_{[a}}\hat{h}\indices{^m_{b]}}\nabla_n\nabla_m\alpha \\ &=T\indices{^n_a_b}\nabla_n\alpha-2t_{[a}\xi^n\nabla_{|n|}\nabla_{b]}\alpha-2t_{[b}\xi^n\nabla_{a]}\nabla_n\alpha,
    \end{align*}
    so that $D_{[a}D_{b]}\alpha=0$ iff $T\indices{^n_a_b}\nabla_n\alpha-2t_{[a}\xi^n\nabla_{|n|}\nabla_{b]}\alpha-2t_{[b}\xi^n\nabla_{a]}\nabla_n\alpha=0$. But
    \begin{align*}
        \xi^r\xi^s(T\indices{^n_r_s}\nabla_n\alpha-2t_{[r}\xi^n\nabla_{|n|}\nabla_{s]}\alpha-2t_{[s}\xi^n\nabla_{r]}\nabla_n\alpha)=0
    \end{align*}
    and 
    \begin{align*}
        h^{ra}\xi^s(T\indices{^n_r_s}\nabla_n\alpha&-2t_{[r}\xi^n\nabla_{|n|}\nabla_{s]}\alpha-2t_{[s}\xi^n\nabla_{r]}\nabla_n\alpha)\\
        &=\xi^sT\indices{^n^a_s}\nabla_n\alpha+h^{ar}\xi^n\nabla_{n}\nabla_{r}\alpha-\xi^nh^{ar}\nabla_{r}\nabla_n\alpha\\
        &=\xi^sT\indices{^n^a_s}\nabla_n\alpha-\xi^nh^{ar}T\indices{^m_r_n}\nabla_m\alpha\\
        &=\xi^sT\indices{^n^a_s}\nabla_n\alpha-\xi^nT\indices{^m^a_n}\nabla_m\alpha\\
        &=0,
    \end{align*}
    so this is equivalent to the requirement that $h^{ra}h^{sb}(T\indices{^n_r_s}\nabla_n\alpha-2t_{[r}\xi^n\nabla_{|n|}\nabla_{s]}\alpha-2t_{[s}\xi^n\nabla_{r]}\nabla_n\alpha)=T^{nab}\nabla_n\alpha=0$ for all $\alpha$. But this will be true just in case $T^{abc}=0$.
\end{proof}

The second proposition is this:
\begin{prop}
    Let $\nabla$ be a non-metric connection on $\langle M, t_a, h^{ab}\rangle$, and let $D$ denote the induced spatial derivative operator on each spacelike hypersurface $S$. Then $D$ is metric compatible iff $Q\indices{^a_b}=Q^{abc}=0$. 
\end{prop}
\begin{proof}
    First, we have
    \begin{equation*}
        D_ah^{bc}=\hat{h}\indices{^n_a}\nabla_nh^{bc}=\hat{h}\indices{^n_a}Q\indices{_n^b^c}=Q\indices{_a^b^c}-t_a\xi^nQ\indices{_n^b^c},
    \end{equation*}
    so that $D_ah^{bc}=0$ iff $Q\indices{_a^b^c}-t_a\xi^nQ\indices{_n^b^c}=0$. Since $\xi^m(Q\indices{_m^b^c}-t_m\xi^nQ\indices{_n^b^c})=0$, this latter condition is equivalent to the requirement that $h^{am}(Q\indices{_m^b^c}-t_m\xi^nQ\indices{_n^b^c})=Q^{abc}=0$.

    Next, we have
    \begin{equation*}
        D_at_b=\hat{h}\indices{^n_a}\hat{h}\indices{^m_b}\nabla_nt_m=\nabla_at_b-\xi^nt_a\nabla_nt_b-\xi^mt_b\nabla_at_m+t_at_b\xi^n\xi^m\nabla_nt_m,
    \end{equation*}
    so that $D_at_b=0$ iff $\nabla_at_b-\xi^nt_a\nabla_nt_b-\xi^mt_b\nabla_at_m+t_at_b\xi^n\xi^m\nabla_nt_m=0$. But
    \begin{align*}
        \xi^r(\nabla_rt_b-\xi^nt_r\nabla_nt_b&-\xi^mt_b\nabla_rt_m+t_rt_b\xi^n\xi^m\nabla_nt_m)\\
        &=\xi^r\nabla_rt_b-\xi^n\nabla_nt_b-\xi^r\xi^mt_b\nabla_rt_m+t_b\xi^n\xi^m\nabla_nt_m\\
        &=0
    \end{align*}
    so this is equivalent to the requirement that $h^{ra}(\nabla_rt_b-\xi^nt_r\nabla_nt_b-\xi^mt_b\nabla_rt_m+t_rt_b\xi^n\xi^m\nabla_nt_m)=\nabla^at_b-\xi^mt_b\nabla^at_m=\hat{h}\indices{^m_b}Q\indices{^a_m}=0$. Since $\hat{h}\indices{^a_b}$ is nowhere vanishing, this will hold just in case $Q\indices{^a_b}=0$.
\end{proof}


\printbibliography

\end{document}